\documentclass[]{iucr}              
\usepackage{bm}
\usepackage{subfig}
\usepackage{graphicx}
\usepackage{wrapfig}
\usepackage{nameref}
\usepackage{blindtext}
\usepackage{mathtools}
\usepackage{amsmath}
\usepackage{wasysym}

\usepackage[hyphens,spaces,obeyspaces]{url}
\usepackage{color}
\usepackage{soul}

     \journalcode{S}              

\begin{document}                  


\title{Modelling phase imperfections in compound refractive lenses} 

\shorttitle{Simulation of real CRL imperfections}

\cauthor[]{Rafael}{Celestre}{rafael.celestre@esrf.eu}{\aff}
\author[]{Sebastien}{Berujon}
\author[]{Thomas}{Roth}
\author[]{Manuel}{Sanchez del Rio}
\author[]{Raymond}{Barrett}

\aff[]{ESRF - The European Synchrotron, 71 Avenue des Martyrs, 38000 Grenoble, \country{France}}


\maketitle                        

\begin{synopsis}
We present coherent- and partially-coherent accurate wavefront propagation simulations using "Synchrotron Radiation Workshop" (SRW) through thick 2D-Be-CRL taking into account the effects of phase errors obtained by X-ray speckle vectorial tracking (XSVT) at the "BM05 - Instrumentation Beamline" at the ESRF.
\end{synopsis}

\begin{abstract} 
We describe a framework based on physical-optics for simulating the effect of imperfect compound refractive lenses (CRL) upon an X-ray beam, taking into account measured phase errors obtained from at-wavelength metrology. We model, with increasing complexity, a CRL stack as a single thin phase element, then as a more realistic compound-element including absorption and thickness effects and finally, we add realistic optical imperfections to the CRL. Coherent and partially-coherent simulations using the "Synchrotron Radiation Workshop" (SRW) are used to evaluate the different models, the effects of the phase errors and to check the validity of the design equations and suitability of the figures of merit.

\end{abstract}


\section{Introduction}\label{section:intro}

The use of refractive optics for the focusing of X-rays dates back to the mid-1990's \cite{Tomie1994, Snigirev1996} which is relatively recent when compared to the use of diffractive- (early 1930's) and reflective-optics (late 1940's). Although a recent development, X-ray lenses are used at most high-energy synchrotron facilities\footnote{Just before the ESRF shut down for the installation of the new ESRF-EBS storage ring, in December 2018, roughly 40$\%$ of the existing beamlines used X-ray lenses in their optical setup.} and X-ray Free-Electron Lasers (XFEL) either for beam conditioning, final focusing of the X-rays into the sample or for imaging. The recent development and establishment of fourth-generation synchrotron light sources - as upgrades of existing machines or the construction of new facilities - and the emergence of the XFEL poses a new challenge for X-ray optics: wavefront preservation, as at modern sources the X-ray beam quality at the sample is primarily limited by the optical quality \cite{Schroer2014, Yabashi2014}. 

Under certain conditions, X-ray lenses are well adapted for situations where minimising wavefront distortions is important \cite{Roth2017, Seiboth2017, Kolodziej2018}. To understand their impact on the optical design of complete beamlines, it is necessary to be able to simulate them realistically. The basic implementation of X-ray lenses is already available on the two most widespread beamline simulation tools: SHADOW \cite{SanchezdelRio2011} and SRW \cite{Chubar1998}. Both implementations, although based on different schemes, ray tracing \cite{Alianelli2007} and wave optics \cite{Baltser2011} respectively, are based on an ideal model combining refraction and absorption for the stacked lenses. Since then, much has been done in terms of refining the modelling of ideal X-ray lenses \cite{Umbach2008, SanchezdelRio2012, Osterhoff2013, Simons2017, Pedersen2018} and, to a certain extent, the modelling of optical imperfections \cite{Pantell2001, Andrejczuk2010, Gasilov2017, Osterhoff2017}. With exception of the work of \citename{Roth2014}~\citeyear{Roth2014}, investigating the inner structure of X-ray lenses, the present models consider only lens shape and  departure from a perfect parabolic shape to either test the limits of figure errors and fabrication defects, or to improve lens shape and focusing quality. These, however, do not include the data from real lenses metrology, as is routinely done for X-ray mirrors simulations \cite{SanchezdelRio2016}. Furthermore, it is important to have simulation tools to allow for the accurate implementation of synchrotron or XFEL light sources, allowing the CRL to be included in a complete beamline configuration in combination with other optical elements. This is possible with both SHADOW and SRW. 

In this work, we propose a framework for simulating CRL taking into account their thickness, absorption and individual lens phase errors measured with at-wavelength metrology. Such phase errors can arise from  material inhomogeneities (voids, impurities) and/or figure errors from the lens forming process.  Our approach is fully compatible with SRW as the X-ray optics community interest shifts to the study of wavefront preservation and tolerancing in low-emittance synchrotron or XFEL beamlines. However, the extension and application of this methodology to ray-tracing \cite{Rebuffi2016} and hybrid methods \cite{Shi2014} is possible. 

This paper is organised as follows: section~2 introduces basic design equations necessary for modelling X-ray lenses and figures of merit to evaluate the optical performance of the CRL. With increasing complexity, section~3 introduces the complex transmission element and from it derives the representation of the CRL and their phase errors used for accurate simulations of real imperfections. Section~4 presents the simulations in two groups: coherent and partially-coherent simulations. This is where the evaluation of the different models, the effects of the lens imperfections and checking the validity of the design equations and suitability of the figures of merit is done. Finally, the results are discussed and the main conclusions are drawn.

\newpage

\section{Compound refractive lenses}\label{section:CRL}

In this section, we recall some important design equations in \S\ref{section:CRL_basics} and figures of merit used when assessing the X-ray focusing quality in \S\ref{section:CRL_FoM}.

\subsection{CRL anatomy}\label{section:CRL_basics}
\noindent

X-ray lenses may have different surface shape: in initial experiments \cite{Snigirev1996} a cylindrical surface was used, which was soon replaced by a parabolic shape that almost completely removes geometrical aberrations \cite{Lengeler1999}. Parabolic lenses are the most used X-ray lenses in CRL as they can focus 1D (cylinder with parabolic section) or in 2D (paraboloid of revolution). It is worth noting, however, that, although less usual, X-ray lenses can assume other shapes: an elliptical profile when focusing collimated beams \cite{Evans-Lutterodt2003}, or a Cartesian oval for point-to-point focusing \cite{SanchezdelRio2012}. However, parabolic shapes always present a very good approximation to geometric focusing and reduce the geometrical aberrations to levels that are smaller than contributions from the fabrication errors and diffraction effects.

\begin{figure}
    \centering
    \includegraphics[width=7cm]{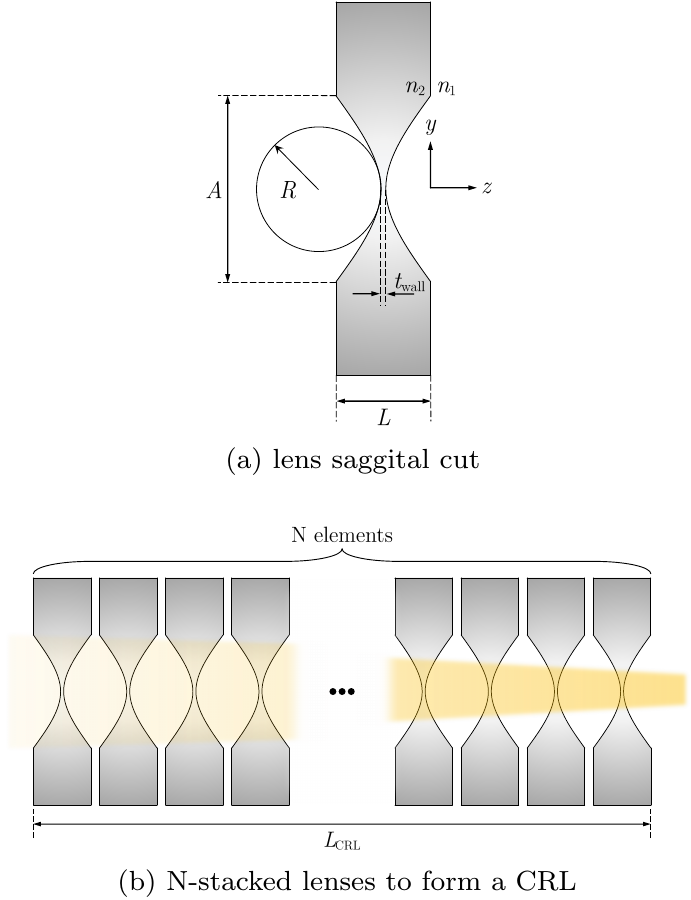} 
    \caption{(a) Sagittal cut of a X-ray lens showing its main geometrical parameters. This concave lens focuses X-rays in the $y$-direction if $n_1>n_2$. (b) N-stacked lenses. A single X-ray lens refracts very weakly. To overcome this drawback - pointed out as early as the late 1940's \cite{Kirkpatrick1948} - lenses are usually stacked, hence "compound" in compound refractive lenses.}
    \label{fig:CRLs}
\end{figure}

Very often X-ray lenses are defined by a small set of parameters as shown in Fig.~\ref{fig:CRLs}(a). These are: \textit{i}) material; \textit{ii}) apex radius of curvature ($R_x$, $R_y$); and \textit{iii}) lens thickness ($L$) or geometrical aperture ($A$) and \textit{iv}) distance between the apices of the parabolas ($\text{t}_{\text{wall}}$).

We begin by defining the optical power $F=f^{-1}$ of a single refracting surface of radius $R$, where $f$ is its focal length. The index of refraction for the X-ray regime can be expressed as a complex number: $n = 1-\delta+i\cdot\beta$, with $\delta$ being the refraction index decrement and $\beta$, the absorption index - both strongly dependent on energy and material. With the X-ray beam moving along the positive $z$-direction on Fig.~\ref{fig:CRLs}, the refracting power of the vacuum/lens interface is given by:
\begin{equation}\label{eq:Focus_simple}
    F_{x,y}\equiv\frac{1}{f_{x,y}}=\frac{n_2-n_1}{-R_{x,y}}=\frac{\delta}{R_{x,y}}.
\end{equation}{}
Here we will consider only the real part of the indices of refraction as this governs the focusing effect of the lenses. As illustrated by Fig.~\ref{fig:CRLs}(a), lenses are typically formed by two refracting surfaces of nominally the same radii. From paraxial optics, the total optical power of refracting surfaces in intimate contact is the sum of their powers. The same is valid for the cases where the distance between them can be ignored. Typical materials used for X-ray lenses have $10^{-7}\leq\delta\leq10^{-3}$ for their usual application energies \cite{Serebrennikov2016}. To overcome the weak refraction of a single element, several X-ray lenses are stacked \cite{Tomie1994, Snigirev1996}. Still, under the assumption of thin elements, we have:
\begin{equation}\label{eq:CRL_classic}
    f_{\text{thin}\text{~CRL}} = \frac{R_{x,y}}{2N\delta},
\end{equation}{}
where the $2N$ comes from stacking $N$ lenslets with two refracting surfaces each, as shown in Fig.~\ref{fig:CRLs}(b). A correction factor can be added to Eq.~\ref{eq:CRL_classic} in order to account for the thick-element nature of the CRL, as proposed by \citename{Kohn2002}~\citeyear{Kohn2002}. The corrected focal length for a thick CRL is given by:
\begin{equation}\label{eq:CRL}
        f_{\text{CRL}} = \frac{R_{x,y}}{2N\delta}+\frac{L_{\text{CRL}}}{6}.
\end{equation}{}
This focal distance is taken from the middle of the CRL and $L_{\text{CRL}}$ is the CRL longitudinal size, that is, distance from the front surface of the first optical element to the back surface of the last lens.

\begin{figure}
    \centering
    \includegraphics[width=7.5cm]{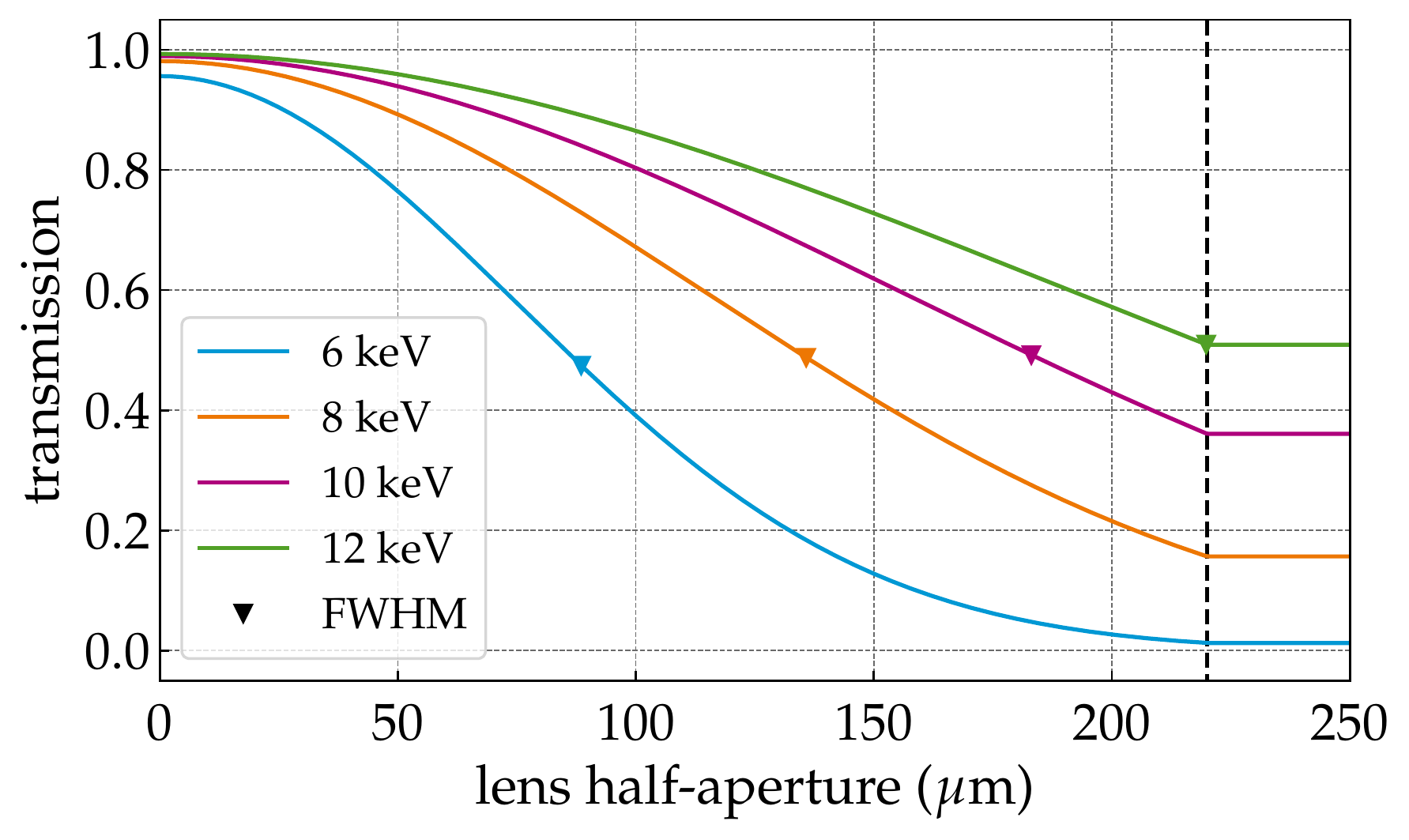} 
    \caption{Intensity transmission profile of a CRL composed of 10 2D-Beryllium lenses with nominal radius $R_{x,y}=50~\mu\text{m}$, geometric aperture $A_{\diameter}=440~\mu\text{m}$ and $t_\text{wall}=10~\mu$m at different photon energies. Vertical dashed line represents the lens geometrical half-aperture. The calculations were done using the "Synchrotron Radiation Workshop" computer code \cite{Chubar1998}.}
    \label{fig:EffectiveAperure}
\end{figure}

Another important parameter for optical design is the lens geometrical aperture $A$, as it provides an upper bound for the numerical aperture of the system and, ultimately, to the theoretical optical resolving power. Assuming a parabolic profile of the refracting surface, the lens geometrical aperture can be calculated as:
\begin{equation}\label{eq:A}
    A_{x,y} = 2\sqrt{(L-\text{t}_\text{wall})R_{x,y}},
\end{equation}{}
where $L$ is the lenslet thickness and $\text{t}_\text{wall}$ is the distance between the apices of the parabolas, commonly referred to as web thickness. 
{}
Due to absorption, the geometrical aperture defined in Eq.~\ref{eq:A} is greater than or equal to the \textit{effective} lens aperture as indicated by \cite{Kohn2017}. There are several reported ways of defining the \textit{effective} lens aperture. Figure~\ref{fig:EffectiveAperure} shows the transmitted intensity profile of a CRL composed of 10 2D-Beryllium lenses with nominal radius $R_{x,y}=50~\mu\text{m}$ and circular geometric aperture $A_{\diameter}=440~\mu\text{m}$ at different energies. Unlike visible optics, where the transmitted intensity profile within the aperture, closely follows that of the illumination, the transmitted profile through a (stack of) X-ray lens(es) has strong absorption towards the edge, which defines the CRL as an apodised optical system.

{}

\subsection{CRL performance}\label{section:CRL_FoM}

Here we present the reader with some other useful figures of merit commonly used for evaluating the performance of optical systems. 

\subsection*{Diffraction limited focal spot}\label{section:PSF}

Even an ideal and aberration-free finite optical element is not able to image a point-source to a point-like image. Limiting the extent of the focusing element by defining an aperture will induce diffraction effects on the wavefront and these will limit the smallest reachable focus spot size. The normalised response of the optical system to this point-like source input is called the point-spread-function (PSF). For a system with circular aperture and uniform amplitude across the exit pupil, the intensity of such focused beam at the image plane is proportional to a squared first-order Bessel function of the first kind (Airy pattern). The FWHM of the central cone is given by:
\begin{align}\label{eq:PSF}
    d = 1.22\lambda (1-M)\frac{f_{\text{CRL}}}{A},
\end{align}{}
where the $M$ is the magnification of the system, which goes to zero for a plane wave or a very distant source. Systems with nonuniform illumination at the pupil exit, in our case apodised systems approaching a Gaussian illumination (see Fig.~\ref{fig:EffectiveAperure}), may present a different PSF shape depending on the truncation imposed by the aperture. A very weakly truncated focusing system (eg. transmission curve for 6~keV in Fig.~\ref{fig:EffectiveAperure}) will have a Gaussian-shaped focal spot as little to no cropping occurs and therefore diffraction effects can be neglected. Increasing the truncation of the beam enhances diffraction effects from the geometric aperture. A strongly truncated focusing system (eg. transmission curve for 12~keV in Fig.~\ref{fig:EffectiveAperure}) will have a PSF that resembles the diffraction pattern in the far-field associated with the aperture of the system\footnote{The far-field diffraction pattern of a circular aperture is a squared first-order Bessel function profile while a square aperture will produce a 2D sinc-squared pattern \cite{Guasti1993}.} \cite{Mahajan1982, Mahajan1986}.

\subsection*{Tolerance conditions for aberrations}\label{section:Strehl}
Introducing errors to the optical system will reduce the peak intensity in the PSF. The ratio between the peak intensities of the aberrated- and non-aberrated PSF of a system with the same aperture and focal length is referred to as the Strehl ratio - cf. \S9.1.3 in \cite{born_wolf1999}. The optical aberrations on the exit pupil of an optical system can be described by the aberration function $\Phi(x,y)$, with the dimension of metres, which represents any deviation in shape from an ideal profile. For small aberration values, the Strehl ratio can be approximated\footnote{Eqs.~\ref{eq:Strehl}-\ref{eq:Mahajan} were obtained using a fully-coherent illumination of the optical system, however, defyning the Strehl ratio as the ratio between the peak intensities of the aberrated- and non-aberrated optical under study transcends the nature of the illumination. A more complete derivation of the Strehl ratio (Eq.~\ref{eq:Strehl}) can be found in \S9.1.3 - \textit{A relation between the intensity and the average deformation of wave-fronts} in \cite{born_wolf1999}.} by:
\begin{equation}\label{eq:Strehl}
    S_{\text{ratio a}}=\frac{I_{\text{aberrated}}}{I_{\text{aberration free}}}\approx1-\bigg(\frac{2\pi}{\lambda}\bigg)^2\Delta\Phi^2,
\end{equation}{}
where $\lambda$ is the wavelength and $\Delta\Phi$ is the standard deviation of the aberration function $\Phi(x,y)$. An important consequence of Eq.~\ref{eq:Strehl} is that the reduction in the peak intensity on the focal plane does not depend on the type of aberration nor the focal length of the optical system, but on its standard deviation across the exit pupil of the optical system \cite{born_wolf1999}. Alternative expressions to Eq.~\ref{eq:Strehl} are available in \S8.3 of \cite{Mahajan2011}, namely:
\begin{align}\label{eq:Marechal}
    S_{\text{ratio b}}\approx\bigg[1-\bigg(\frac{2\pi}{\lambda}\bigg)^2\frac{\Delta\Phi^2}{2}\bigg]^2,
\end{align}
known as the Mar\'echal expression and:
\begin{align}\label{eq:Mahajan}
    S_{\text{ratio c}}\approx\exp{\bigg[- \bigg(\frac{2\pi}{\lambda}\bigg)^2\Delta\Phi^2\bigg]},
\end{align}
an empiric expression that fits better numerical results \cite{Wetherell1980}. However, for strong aberrations, there is no simple analytic expression to describe the relation between the Strehl ratio and the standard deviation of the aberration function $\Phi(x,y)$ \cite{Kessler81}. 

It is possible to define an arbitrary minimum acceptable value to the Strehl ratio when evaluating an optical element quality (tolerancing). This value depends on the final application and the desired performance. However, a value of $S_\text{ratio}\geq0.8$ is commonly found throughout literature as an indicator of a well-corrected optical system\footnote{This comes from historic reasons: both Rayleigh's $\lambda/4$ criterion for spherical aberrations (1879) and the extended Mar\'echal criterion for optical quality (1943) yield in a Strehl ratio of $\sim0.8$ \cite{born_wolf1999}.}. Inserting $S_\text{ratio}\geq0.8$ in Eq.~\ref{eq:Strehl}, one obtains:
\begin{equation}\label{eq:MarechalCriterion}
    |\Delta\Phi|\leq\frac{\lambda}{14},
\end{equation}{}
which is known as the Mar\'echal criterion for optical quality. Equations~\ref{eq:Marechal}~and~\ref{eq:Mahajan} give similar limits: $\lambda/13.67$ and $\lambda/13.30$, respectively. In order to apply  Eq.~\ref{eq:MarechalCriterion} to the case of an X-ray lens, we use Eq.~\ref{eq:aux_funcs_transb} (from \S\ref{section:TransmissionElement}$-$\nameref{section:TransmissionElement}) with $\Delta\phi =\frac{2\pi}{\lambda}\delta\sigma_z=\frac{2\pi}{\lambda}|\Delta\Phi|$, where $\Delta\phi$ is the  standard deviation of the phase, and we replace the projected thickness $\Delta z$ with the standard deviation of the projected figure error $\sigma_z$:
\begin{align}\label{eq:ThickLim}
    \sigma_z &\leq \frac{\lambda}{14\delta}.
\end{align}{}
Equation \ref{eq:ThickLim} gives an upper limit to the standard deviation of accumulated figure errors for X-ray lenses in order to comply with the Mar\'echal criterion of tolerable wavefront aberrations, or in other words, to sustain a $S_\text{ratio}\geq0.8$. For a more complete discussion on the aberrated PSF, Strehl ratio and tolerance conditions for primary aberrations, refer to \S9 from \cite{born_wolf1999} and \S8 from \cite{Mahajan2011}.

\subsection{Chromatic aberrations}\label{section:CRL_EnergyDep}

The optical properties of the X-ray lenses are strongly dependent on the wavelength as both $\delta$ and $\beta$ have an energy dependency. This causes chromatic aberrations and limitations on the optical performance of the CRL under an X-ray beam with finite bandwidth. The chromaticity of X-ray lenses can be used favourably for X-ray harmonic rejection from insertion devices and coarse X-ray spectrum filtering \cite{Vaughan2011, Polikarpov2014}.

\section{CRL: physical optics modelling}\label{section:Modelling}

In this section, we present the models for accurately representing a realistic CRL following wave-optics representation \cite{Goodman2017}. We start by defining a complex transmission element and, with increasing complexity, we present the different models that are based on this complex transmission element concept. 

\subsection{The complex transmission element}\label{section:TransmissionElement}

The amplitude transmission of radiation through matter can be expressed as a complex operator - \S2 in \cite{Paganin2006}: 
  \begin{align}\label{eq:complex_transmission}
\textbf{T}(x,y,z;\lambda) &=\exp{\bigg\{\frac{-2\pi i}{\lambda}\int_C\big[\delta(x,y,z;\lambda)-i\cdot\beta(x,y,z;\lambda)\big]\text{d}s\bigg\}},
\end{align}{}
with $z$ being along the beam direction, $x$ and $y$ are the transverse coordinates to $z$, $\lambda$ is the wavelength and  $\int_C$ is a path integral along $\text{d}s$. Favouring a more compact notation, we drop here the explicit energy dependency of the index of refraction. The $z-$dependence of $\delta$ and $\beta$ is abandoned in the projection approximation, which is often valid when modelling refractive optics\footnote{Multi-slicing (MS) techniques are often used to deal with the cases where, for weak scatterers, $n(x,y,z)\neq n(x,y,z+\Delta z)$. Here $\Delta z$ is an incremental distance along the propagation direction. Chapters \S2.2 and \S2.7 in \cite{Paganin2006} deal in more details with the projection approximation and the MS technique, which was first described in the context of the scattering of electrons by atoms and crystals \cite{Cowley1957}.}. The integral in Eq.~\ref{eq:complex_transmission} reduces to:
\begin{align}\label{eq:transmission}
\textbf{T}(x,y,z) &=\exp{\bigg\{\frac{-2\pi i}{\lambda}\int_0^{\Delta z}\big[\delta(x,y)-i\cdot\beta(x,y)\big]\text{d}s\bigg\}}\nonumber\\
&=\exp{\bigg\{\frac{-2\pi i}{\lambda} \big[\delta(x,y)-i\cdot\beta(x,y)\big]\Delta z \bigg\}}\nonumber\\
\textbf{T}(\Delta z) &=\sqrt{\text{T}_\text{BL}}\exp{\big(i\phi\big)},
\end{align}{}
where:
\begin{subequations}
\begin{align}   
    \text{T}_{\text{BL}}&=\exp{\bigg(-\frac{4\pi}{\lambda}\beta\Delta z\bigg)}\label{eq:aux_funcs_transa}  \\
    &=\exp{\big(-\mu\Delta z\big)},\nonumber\\
    \phi&=-\frac{2\pi}{\lambda}\delta\Delta z.\label{eq:aux_funcs_transb}
\end{align}
\end{subequations}

The integration path $\text{d}s$ is along the $z-$direction. It is proportional to the projected thickness $\Delta z$, which in turn, depends on the transverse coordinates: $\Delta z \equiv \Delta z(x,y)$. Equation~\ref{eq:aux_funcs_transa} shows the absorption experienced by the wavefront when passing through matter (Beer-Lambert law) and Eq.~\ref{eq:aux_funcs_transb} shows the phase-shift experienced by the wavefront. The coefficient multiplying $\Delta z$ in $\text{T}_{\text{BL}}$ is know as linear attenuation coefficient $\mu$. The transmitted electric field is obtained by multiplication of the input field with the complex transmission operator in Eq.~\ref{eq:transmission}, that is, $\textbf{E}_2 = \textbf{T}[\Delta z(x,y)]\cdot\textbf{E}_1$.

\subsection{Ideal thin lens and single lens equivalent}\label{section:SingleLensEq}

\begin{figure}
    \centering
    \includegraphics[width=6cm]{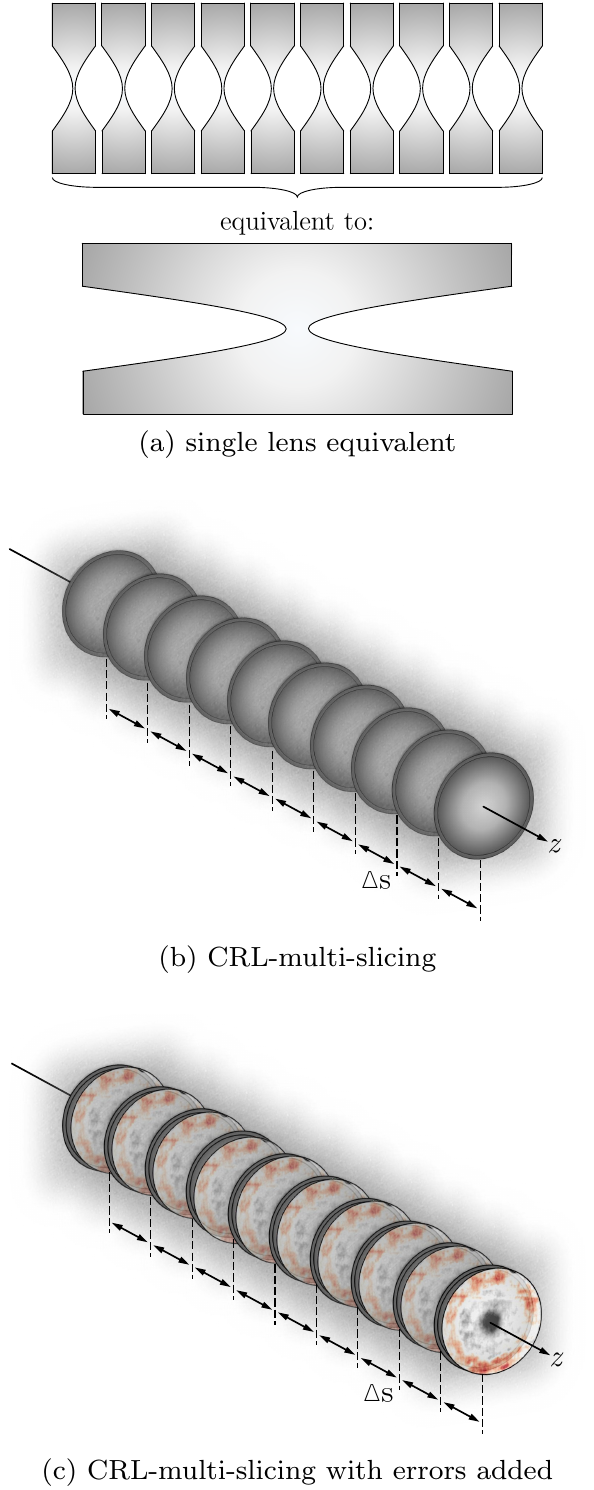} 
    \caption{Hierarchical depiction of the CRL. (a) illustrates a single thin element equivalent of several lenslets. This representation accounts for net refraction and absorption in one transmission element but ignores intra-lens spacing. (b) multi-slice \cite{Munro2019} representation of a CRL. Here each lens of the stack is represented individually by one transmission element. Those are separated by a drift space corresponding to the typical distance between elements ($\Delta\text{s}$). (c) Not only can the CRL be represented as a series of thin elements separated by drift spaces, but also figure errors can be added. They are placed directly after the thin element representing a single X-ray lens.}
    \label{fig:models}
\end{figure}

At any point inside the geometric aperture of a single bi-concave paraboloidal X-ray lens, the projected thickness $\Delta z$ can be calculated as:
\begin{equation}\label{eq:ProjecThick}
    \Delta z(x,y) = \frac{x^2}{R_x}+\frac{y^2}{R_y}+\text{t}_\text{wall}.
\end{equation}{}
Eq.~\ref{eq:ProjecThick} can be substituted into Eqs.~\ref{eq:aux_funcs_transa} and \ref{eq:aux_funcs_transb} to retrieve the complex transmission element expression for an X-ray lens:
\begin{eqnarray}\label{eq:TE_singlelens} 
    \textbf{T}_{\text{single lens}}(\Delta z)~\bullet = \exp{\bigg[i\frac{2\pi}{\lambda}\beta \bigg(\frac{x^2}{R_x}+\frac{y^2}{R_y}+\text{t}_\text{wall}\bigg)\bigg]}\times~~~~~\nonumber\\*
    ~~~~~\times\exp{\bigg[-i\frac{2\pi}{\lambda}\delta\bigg(\frac{x^2}{R_x} + \frac{y^2}{R_y}\bigg)\bigg]} \bullet.
\end{eqnarray}{}
The $\bullet$ symbol represents an arbitrary input electric field. Eq.~\ref{eq:TE_singlelens}, the single lens model, accounts for the absorption (first exponential) and phase shift (second exponential ignoring the constant phase shift induced by $t_\text{wall}$). The complex transmission representing a CRL composed of $N$ elements is, thus, represented by:
\begin{equation}\label{eq:TE_CRL}
    \textbf{T}_{\text{CRL}}(\Delta z)~\bullet = \big[\textbf{T}_{\text{single lens}}(\Delta z)\big]^{N} \bullet.
\end{equation}{}
The model represented by Eq.~\ref{eq:TE_CRL} will be referred to as the single lens equivalent. This model represents a lens stack by a single transmission element with equivalent focal distance and the projected thickness of all the $N$ single lenses as shown in Fig.~\ref{fig:models}(a). 
{}

\subsection{Multi-slicing representation of a CRL}\label{section:MS}

For a CRL composed of a very high number of lenslets, the single-lens equivalent approximation (Eq.~\ref{eq:TE_CRL}) may not be adequate to correctly represent such optical systems mainly due to the thick nature of the stack  - evidenced by Eq.~\ref{eq:CRL}; and due to the progressive focusing inside the CRL \cite{Schroer2005} - exaggerated in Fig.~\ref{fig:CRLs}(b). For such cases, it is possible to adapt the multi-slicing (MS) techniques for the calculation of the transmission of a wavefront through a CRL. Unlike the methods described by \citename{Paganin2006}~\citeyear{Paganin2006} and most recently, by \citename{Li2017}~\citeyear{Li2017} and \citename{Munro2019}~\citeyear{Munro2019}, where a single weakly-scattering optical element is sliced into several slabs, it is sufficient for most practical cases to break down a CRL into its lenses as shown in  Fig.~\ref{fig:models}(b). This can be justified by the fact that at their typical operation energy, the materials used for lens manufacturing have a very low $\delta$ \cite{Serebrennikov2016}, rendering the individual lenslets a weak focusing element where the projection approximation holds \cite{Protopopov1998}. The complex transmission representation of a CRL based on the MS approach is given by:

\begin{equation}\label{eq:TE_CRL_MS}
    \textbf{T}_{\text{CRL-MS}}(\Delta z)~\bullet = \textbf{T}_{\text{single lens}}(\Delta z)\cdot\big[\mathcal{D}({\Delta}\text{s})\cdot\textbf{T}_{\text{single lens}}(\Delta z)\big]^{N-1}\bullet,
\end{equation}{}
where $\mathcal{D}({\Delta}\text{s})$ is the operator formulation of the Fresnel free-space propagation over a distance $\Delta\text{s}$ (distance between the centre of two adjacent lenses), from \S1.4.1 in \cite{Paganin2006}. 

Eq.~\ref{eq:TE_CRL_MS} represents a wavefront $\bullet$ modified by a single lens complex transmission $\textbf{T}_{\text{single lens}}$, followed by free-space propagation $\mathcal{D}({\Delta}\text{s})$ over a distance $\Delta \text{s}$. The multiplication of the resulting electric field by the transmission element and subsequent free-space propagation is done $(N-1)$ times until the $N^{\text{th}}$ lens is reached and the last element of the lens stack is accounted for.

Optical imperfections measured with high spatial resolution can be readily converted into a transmission element by direct application of Eq.~\ref{eq:transmission} to the height profile, provided it is a 2D map of the phase defects. In this case, the height profile will be the projected thickness of $\Delta z(x,y)$ in the preceding equations. The MS model introduced earlier in this section can then be adapted to account for the phase errors of the individual lenses:
\begin{equation}\label{eq:TE_CRL_MS_ERR}
    \textbf{T}_{\text{CRL-MS}}(\Delta z)~\bullet = \textbf{T}_{\text{imperfect lens}}(\Delta z)\cdot\big[\mathcal{D}({\Delta}\text{s})\cdot\textbf{T}_{\text{imperfect lens}}(\Delta z)\big]^{N-1}\bullet,
\end{equation}{}
with:
\begin{equation}
    \textbf{T}_{\text{imperfect lens}}(\Delta z) = \textbf{T}_{\text{figure errors}}(\Delta z)\cdot\textbf{T}_{\text{single lens}}(\Delta z).
\end{equation}{}
This extended version of the MS model is shown in Fig.~\ref{fig:models}(c).

\section{ Analysis of figure errors from metrology with X-ray speckle tracking}\label{section:Figure_errors}

To simulate the CRL performance on a beamline more accurately, one needs a 2D map of surface imperfections. Optical and tactile metrology methods are not the most appropriate to characterise single X-ray lenses due to their geometry \cite{Lyatun2015} and their general insensitivity to subsurface defects (voids, inclusions etc). At-wavelength metrology is often more appropriate for obtaining the figure errors. Several methods are available using X-rays: tomography \cite{Narikovich2017}, grating interferometry \cite{Rutishauser2011}, speckle tracking \cite{Berujon2013} and ptychography \cite{Seiboth2016}.

Each error profile used in the following simulations comes from a real 2D-Be lens individually characterised using X-ray speckle vectorial tracking (XSVT) \cite{Berujon2019b, Berujon2019a} at the \textit{BM05 - Instrumentation Beamline} at the ESRF \cite{BM05}. This technique allows for a high spatial resolution (pixel size of $\sim0.65\times0.65~\mu\text{m}^2$) characterisation of the lens figure error in projection approximation, which can be incorporated directly in the simulations.

\begin{table}[]
\caption{Summary of the main parameters from the metrology of the Be lenses used in the simulations. The bottom rows display the accumulated figure errors calculated by propagating a plane wave through the system; the accumulated figure errors weighted with the system transmission at 8~keV; and the quadrature summation of the individual r.m.s. values from L01 to L10.}
\centering
\label{tab:LensSpecs}
\begin{tabular}{cccccc}
\textbf{lens}      & \textbf{radius}  & \multicolumn{3}{c}{figure errors (r.m.s) $\mu$m}    & \textbf{useful aperture}\\ \cline{3-5} 
\textbf{number}    & $\mu$m           & \textbf{FF} & \textbf{LF}  & \textbf{HF}   & $\mu$m\\ \cline{1-6}
L01              & 49.43            & 0.56        & 0.47       & 0.30        & 425.9\\
L02              & 48.66            & 1.12        & 1.03       & 0.42        & 421.0\\
L03              & 49.18            & 0.90        & 0.74       & 0.52        & 430.9\\
L04              & 49.88            & 2.24        & 2.18       & 0.49        & 427.2\\
L05              & 48.66            & 1.19        & 1.07       & 0.52        & 424.7\\
L06              & 49.26            & 1.15        & 0.85       & 0.77        & 428.4\\
L07              & 49.29            & 0.75        & 0.60       & 0.46        & 433.4\\
L08              & 49.41            & 1.28        & 1.13       & 0.69        & 432.1\\
L09              & 48.71            & 1.43        & 1.36       & 0.44        & 433.3\\
L10              & 48.63            & 0.82        & 0.73       & 0.37        & 417.3\\
\cline{1-6}
\multicolumn{2}{r}{\textbf{accumulated}:}      & 5.22      & 4.91       & 1.77        & 417.3\\
\multicolumn{2}{r}{\textbf{weighted}:}         & 3.64      & 3.54       & 0.85        & $-$\\
\multicolumn{2}{r}{\textbf{quadrature-sum}:}   & 3.88      & 3.53       & 1.63        & $-$\\
\end{tabular}
\end{table}

\subsection*{Aberrations from metrology data}

\begin{figure}
    \centering
    \includegraphics[width=9cm]{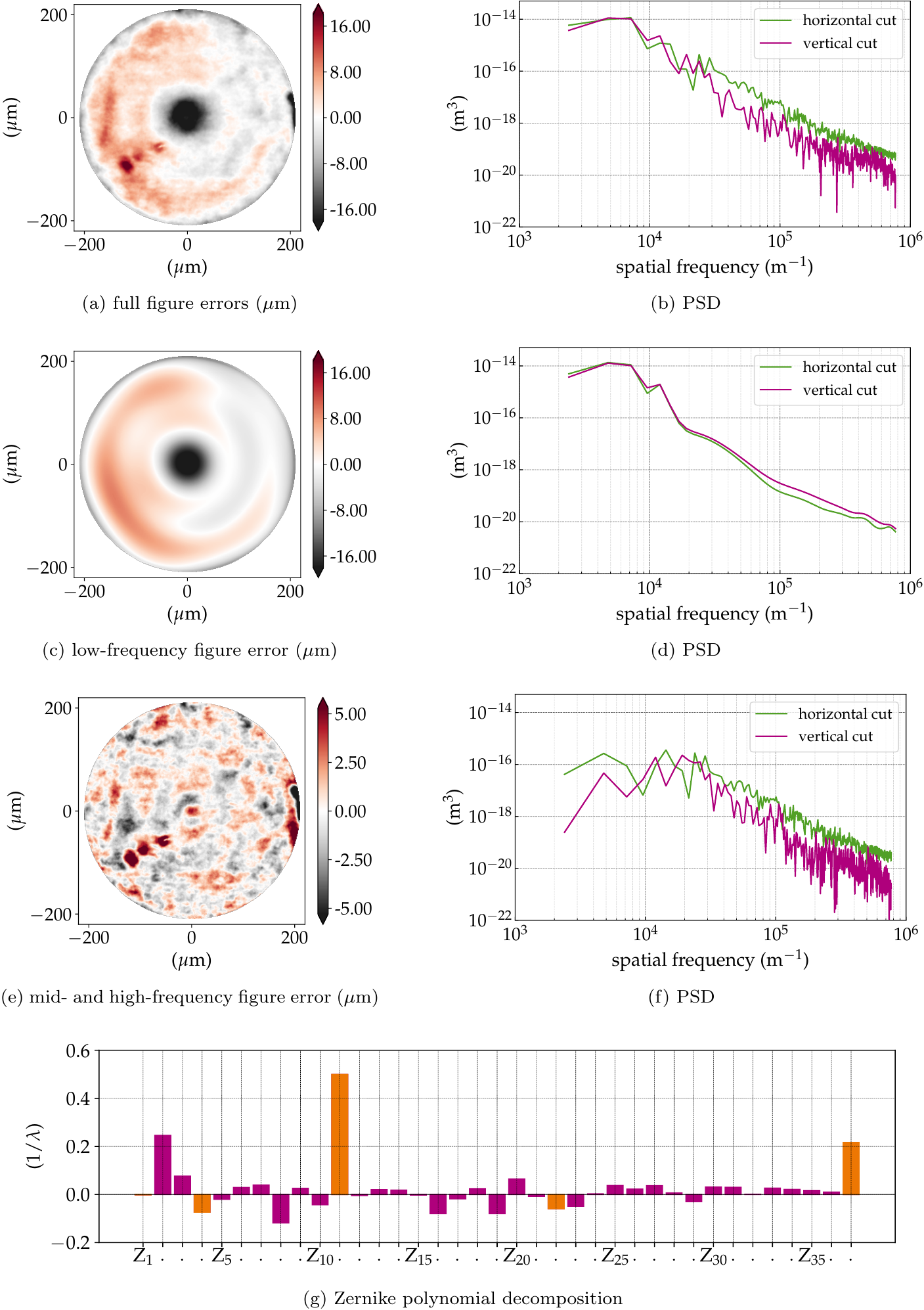} 
    \caption{Accumulated CRL figure errors and their power spectrum density (PSD) for: (a) and (b) full-frequency figure errors (FF); (c) and (d): the low-spatial-frequencies (LF); (e) and (g): mid- and high-spatial frequencies (HF). (g) Zernike polynomials (in Noll notation) amplitude from the transmitted wavefront phase from CRL with full errors. The amplitudes are normalised to the wavefront in \r{A}ngstr\"{o}m. The orange bars indicate rotationally-symmetric terms. Notable contributions are $Z_2$ and $Z_3$ (tilts), $Z_4$ (defocus), $Z_8$ (horizontal coma), $Z_{11}$ (first order spherical aberration) and $Z_{37}$ (third order spherical aberration).}
    \label{fig:phase}
\end{figure}

Following \citename{Harvey1995a}~\citeyear{Harvey1995a}, the figure errors of the lenses can be specified in terms of their spatial frequency, as they often have different effects on the image quality. Three regions are commonly used for that: low-, mid- and high-spatial-frequencies. The low-spatial frequencies are responsible for changing the beam profile and reducing the peak intensity. Mid- and high-spatial frequencies (HF) are responsible for scattering the light around the (focused) beam and have potential for broadening it, together with the expected reduction of the Strehl ratio. The low-frequencies are related to the conventional optical aberrations \cite{born_wolf1999} and they can be described by a set of orthonormal polynomials. For optical systems with a circular aperture, 2D Zernike polynomials are often used \cite{Mahajan2007}, while for rectangular apertures (typical of 1D-focusing lenses) 2D Legendre polynomials are more common \cite{Ye2014}\footnote{Using 2D Zernike and Legendre polynomials for describing conventional optical aberrations in X-ray lenses was first presented by \cite{Koch2016}.}. The mid- and high- frequencies are the residuals from the polynomial fit of the aberrated profile. When referring to the full frequency extent, that is, the addition of the low-, mid- and high-spatial-frequencies, we use "full profile"~(FF). From the analysis of the experimental data we can infer that the low-frequencies span from $\sim500~\mu$m or $2\times10^{3}~\text{m}^{-1}$ (geometrical aperture of a lenslet) to $\sim50~\mu$m or $2\times10^{4}~\text{m}^{-1}$ (power spectral density cut-off - cf. Fig.~\ref{fig:phase}), while the mid- and high-frequencies span from $\sim50~\mu$m or $2\times10^{4}~\text{m}^{-1}$ to $\sim1.3~\mu$m or $0.8\times10^{6}~\text{m}^{-1}$ (obtained from the Nyquist frequency of the measured data: determined by one over twice the detector resolution).
 
Table~\ref{tab:LensSpecs} presents the radius of curvature, r.m.s. value of the figure errors and useful aperture obtained by XSVT for each simulated lenslet and the accumulated error profile, the net erroation seen by a plane wavefront passing through the lens stack. Figure~\ref{fig:phase} presents the accumulated figure errors for the simulated stack, along with their power spectrum density for the full profile, low-, and mid- and high-spatial frequencies; as well as the Zernike polynomial fit of the full profile, which is dominated by primary spherical aberration ($Z_{11}$), tertiary spherical aberration ($Z_{37}$) and horizontal coma ($Z_{8}$). Tilts and defocus ($Z_{2}$, $Z_{3}$ and $Z_{4}$) are not treated here as optical errors.

{}

\section{Simulation results}\label{section:Simulations}

In this section, we present the simulations of the main CRL models. All simulations presented here were done using the "Synchrotron Radiation Workshop" (SRW) \cite{Chubar1998}\footnote{Available at \url{https://github.com/ochubar/srw}}, as it conveniently offers the possibility of fully- and partially-coherent calculations, and presents native parallelisation with the MPI standard \cite{Chubar2011}. Fully coherent calculations were done using a single CPU of an Intel(R) Xeon(R) CPU E5-2680 v4 @ 2.40GHz, while partial coherent simulations used 28 CPUs of the same computer infrastructure (NICE OAR cluster at the ESRF).

\subsection*{Lenses and lens stack}

The simulated lenses are 2D-Be lenses with a nominal radius of $R_{x,y}=50~\mu\text{m}$ chosen as representative of lenses used widely at beamlines at many synchrotrons. Such lenses have typically 1~mm thickness and are held in a 2~mm thick lens frame. If the wall thickness is $\sim30~\text{to}~40~\mu\text{m}$, the lens geometric aperture calculated by applying Eq.~\ref{eq:A}, which gives $A_{\diameter}\le440 \mu\text{m}$. The lens stack is composed of ten lenses piled without any spacing other than the intrinsic lens frame thickness. The transmission through this CRL can be seen in Fig.~\ref{fig:EffectiveAperure}. At 8~keV, the energy used for the simulations, the index of refraction for Beryllium is $n=1-5.318\times10^{-6}+i\cdot2.071\times10^{-9}$. Applying Eq.~\ref{eq:CRL_classic} with $N=1$, one obtains the focal length for a single lens: $f_{\text{lens}}=4.701$~m. The lens stack focal distance can be obtained by applying Eq.~\ref{eq:CRL} with $N=10$ and $L=(N-1)\cdot2~\text{mm}=18~\text{mm}$: $f_{CRL}=473~{\text{mm}}$, giving a magnification of approximately $126:1$ ($M\approx8\times10^{-3}$ for a source 60~m away from the CRL) and a diffraction-limited spot size (Eq.~\ref{eq:PSF}) of $\sim200~\text{nm}$. 

\subsection{Simulations with a coherent wavefront}\label{section:coeh}

For this set of simulations, we used a collimated (plane) wavefront as any deviations from a constant phase and intensity on the exit of the optical system can be immediately attributed to the CRL model being studied.

\subsection*{The PSF: ideal focusing}\label{section:coeh_PSF}

After passage through the CRL model being studied, a plane wave will develop a quadratic phase term that has a curvature radius equivalent to the effective focal distance of the optical system. Table~\ref{tab:FocalLength} compares the calculated focal lengths (Eq.~\ref{eq:CRL_classic} and Eq.~\ref{eq:CRL}) against the focal length extracted from the simulation models. While the single lens equivalent model and the CRL multi-slicing values were obtained using the nominal radius $R=50~\mu$m, the focal length the CRL-MS+FF model was obtained considering the radii from Table~\ref{tab:FocalLength}.

The propagation of the wavefront from the exit pupil of the CRL to the focal length distance (image plane) is equivalent to an optical 2D-Fourier transform of the system pupil function. The PSF of the optical system corresponds to the squared modulus of this Fourier transform, which is the wavefront intensity at the focal plane, considering a plane wave illumination as in \S2.3.1 and \S6.2 from \cite{Goodman2017}. The phase of the propagated field at the focal position along with the normalised PSF for the multi-slice CRL models (without and with the addition of figure errors) can be found in Fig.~\ref{fig:simulations_ideal}(d)-(e), Fig.~\ref{fig:simulations_HF}(d)-(e), Fig.~\ref{fig:simulations_LF}(d)-(e) and Fig.~\ref{fig:simulations_FF}(d)-(e). The calculated FWHM of the central lobe of the PSF for the single-lens equivalent, multi-slicing and multi-slicing with figure errors is displayed on Table~\ref{tab:beamsizes} along with the theoretical diffraction-limited spot size (Eq.~\ref{eq:PSF}). The relative intensities of the aberrated PSF normalised to the ideal case (Strehl ratio) are compiled in Table~\ref{tab:Strehl} and shown in Fig~\ref{fig:MEsimulations}(a).

\begin{table}[]
\label{tab:FocalLength}
\caption{Comparison between theoretical and simulated focal lengths for different CRL models.}
\centering
\begin{tabular}{rccc}
\multicolumn{1}{c}{} & \multicolumn{3}{c}{focal length (m)}\\ \cline{2-4}
\multicolumn{1}{c}{\textbf{lens model}} & \textbf{calculated} &\textbf{fit} &\textbf{difference}\\\cline{1-4}
single lens equivalent   &0.470 (Eq.~\ref{eq:CRL_classic})   &0.470     & -                     \\
CRL multi-slicing        &0.473 (Eq.~\ref{eq:CRL})           &0.473     &$<$0.1\%   \\
CRL-MS + FF              &0.465 (Eq.~\ref{eq:CRL})           &0.465     &$<$0.1\%  \\
\end{tabular}
\end{table}

\begin{table}[]
\label{tab:beamsizes}
\caption{Summary of the beam sizes in FWHM for various CRL models. The extended source image sizes are taken from the partially coherent simulations averaging the intensity of 10$^{4}$ wavefronts.}
\centering
\begin{tabular}{rccc}
&\textbf{PSF}   &\multicolumn{2}{c}{\textbf{extended source image}}\\
\textbf{lens model}     &(nm) &horizontal (nm) &vertical (nm)\\ \cline{1-4}
analytic equations      & 199.8 &605.6 &204.1 \\
single lens equivalent  & 201.7 &598.5 &207.2\\
CRL multi-slicing       & 203.0 &602.4 &208.0\\
CRL-MS+HF               & 202.5 &607.7 &207.0\\
CRL-MS+LF               & 197.6 &640.6 &207.3\\
CRL-MS+FF               & 197.2 &631.9 &209.6\\
\end{tabular}
\end{table}{}

\subsection*{Beam caustics}\label{section:coeh_caustics}

\begin{figure}
    \centering
    \includegraphics[width=15cm]{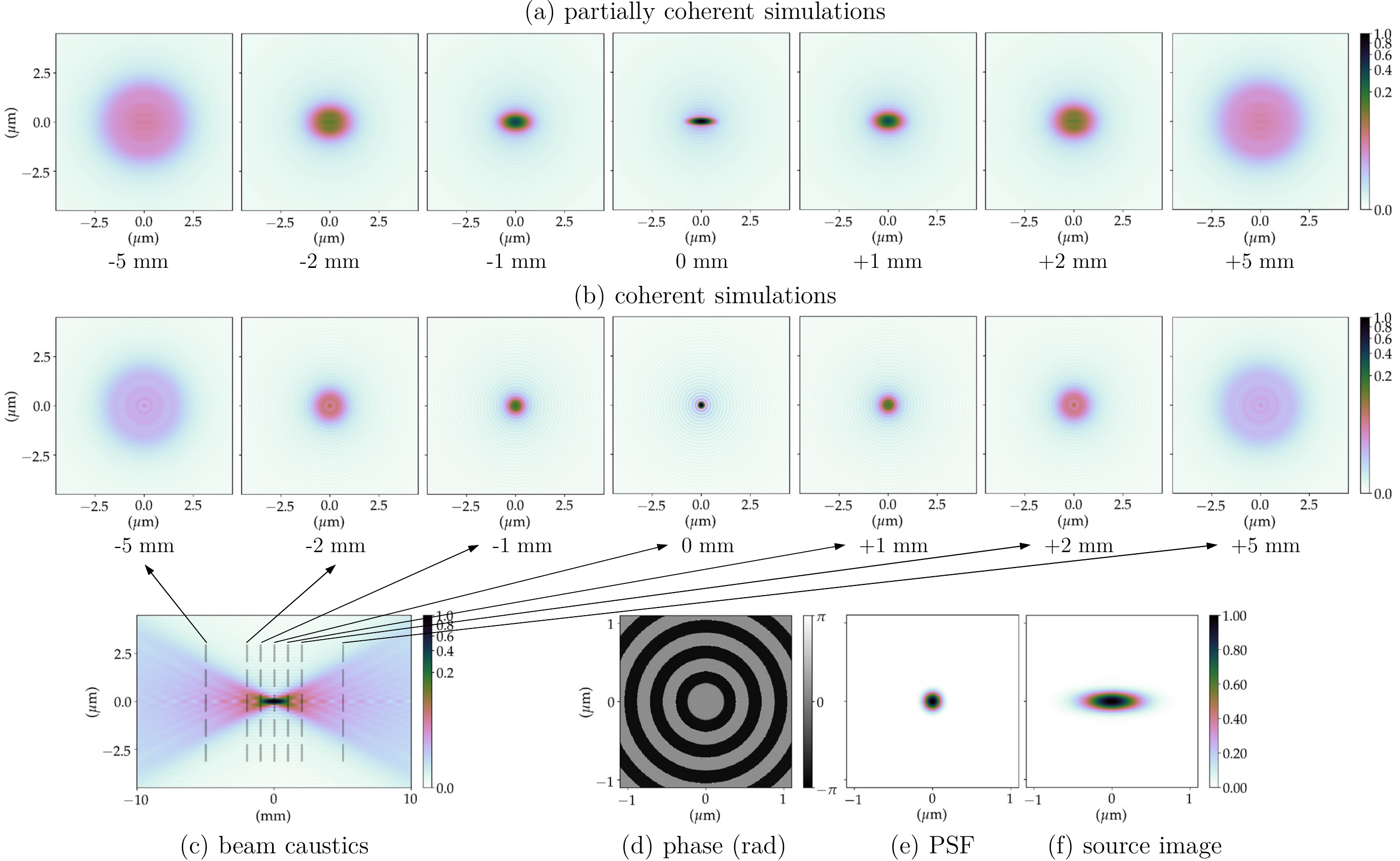} 
    \caption{Ideal CRL-MS model at 8~keV. (a) partially-coherent simulations show the beam profile up- and downstream the focal position averaging 10$^{4}$ wavefronts to simulate the radiation emitted by an undulator; (b) the coherent simulations show the beam profile of a plane wavefront being focused; (c) beam propagation near the focal position (beam caustics) for a fully coherent beam (horizontal cut around $y=0$); (d) phase and (e) intensity of the PSF calculated focusing a plane-wavefront; (f) demagnified image of the undulator photon-source (extended source). }
    \label{fig:simulations_ideal}
\end{figure}

\begin{figure}
    \centering
    \includegraphics[width=15cm]{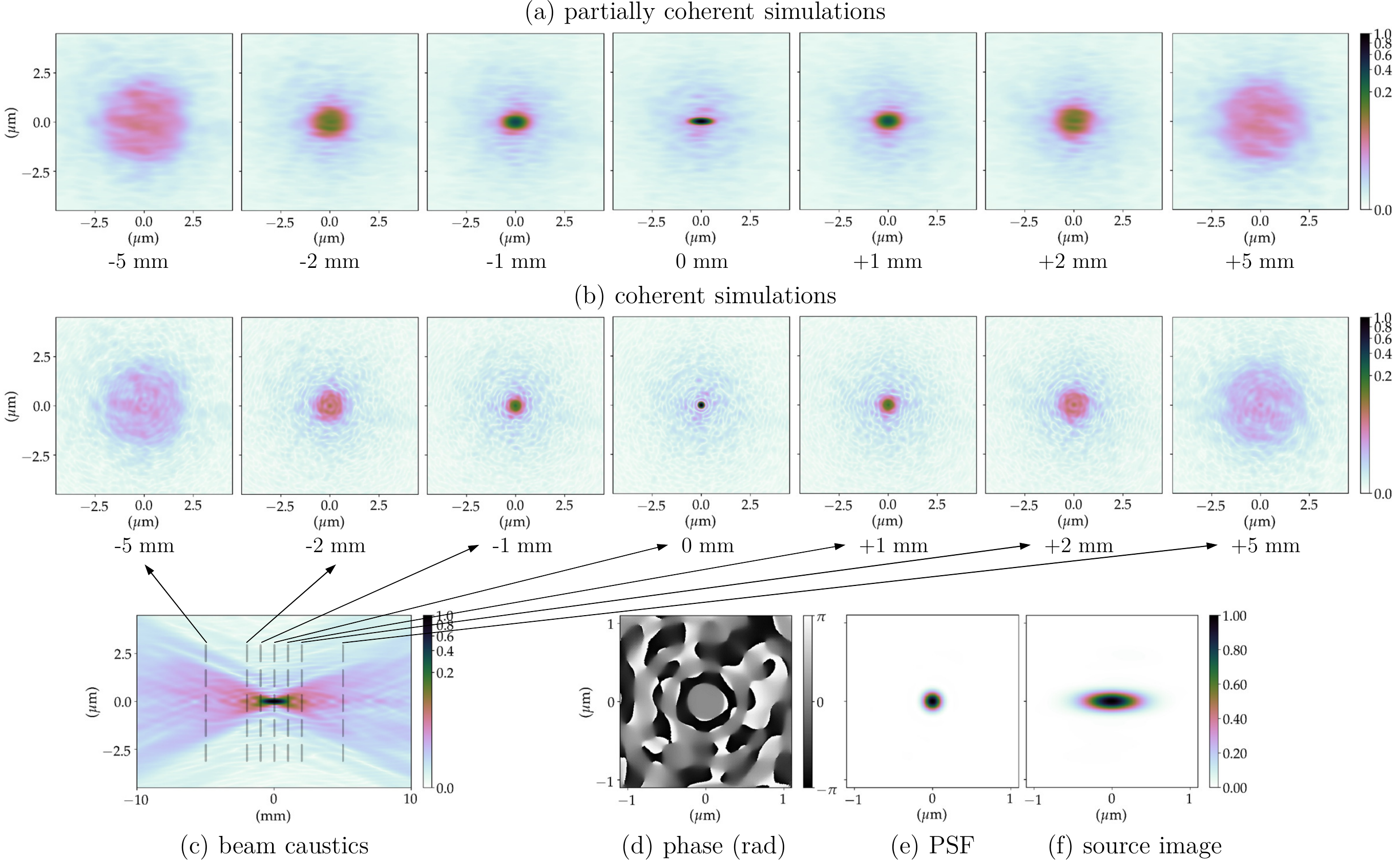} 
    \caption{CRL-MS model with high-frequency figure errors at 8~keV. (a) partially-coherent simulations show the beam profile up- and downstream the focal position averaging 10$^{4}$ wavefronts to simulate the radiation emitted by an undulator; (b) the coherent simulations show the beam profile of a plane wavefront being focused; (c) beam propagation near the focal position (beam caustics) for a fully coherent beam (horizontal cut around $y=0$); (d) phase and (e) intensity of the PSF calculated focusing a plane-wavefront; (f) demagnified image of the undulator photon-source (extended source).}
    \label{fig:simulations_HF}
\end{figure}

\begin{figure}
    \centering
    \includegraphics[width=15cm]{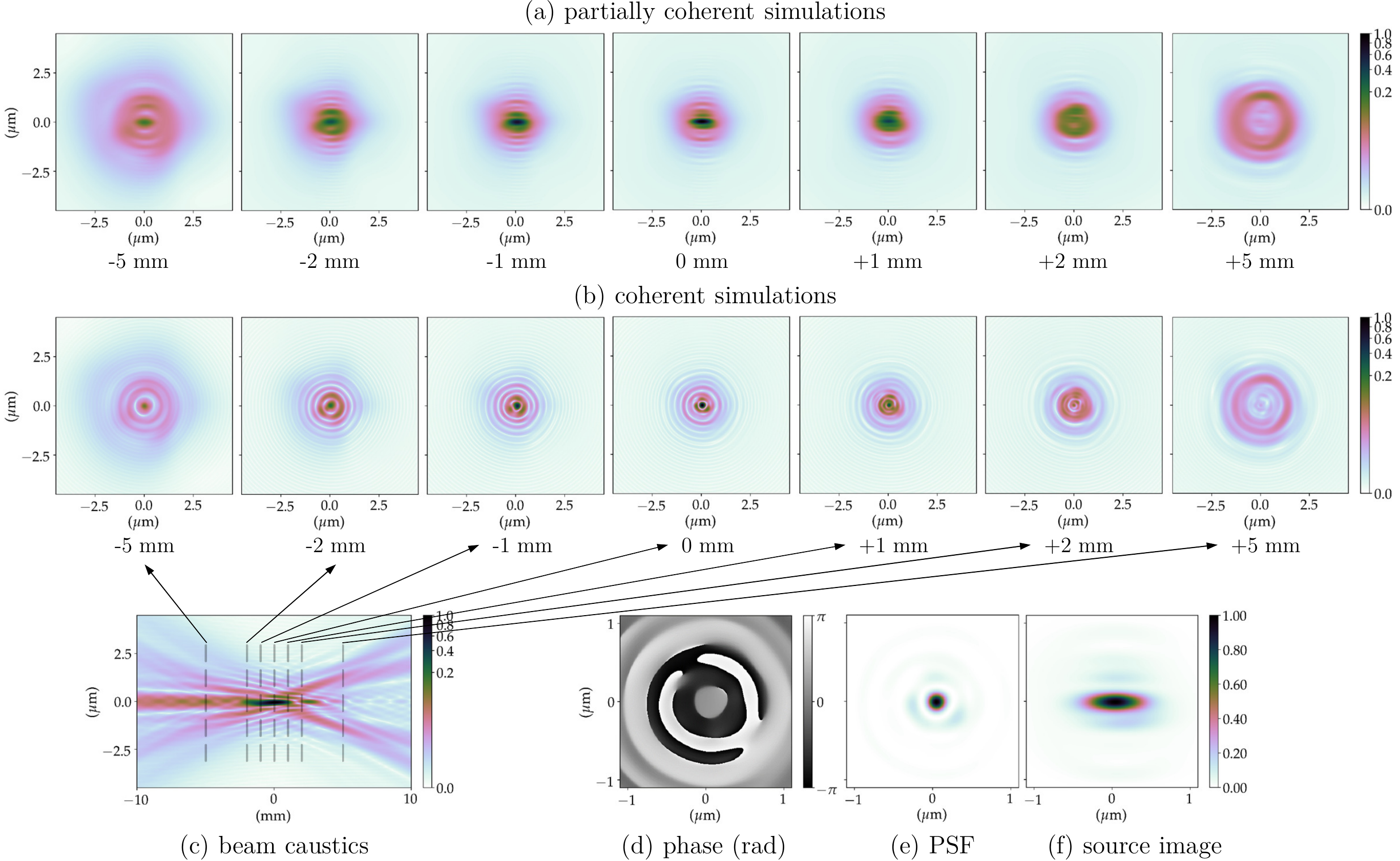} 
    \caption{CRL-MS model with low-frequency figure errors at 8~keV. (a) partially-coherent simulations show the beam profile up- and downstream the focal position averaging 10$^{4}$ wavefronts to simulate the radiation emitted by an undulator; (b) the coherent simulations show the beam profile of a plane wavefront being focused; (c) beam propagation near the focal position (beam caustics) for a fully coherent beam (horizontal cut around $y=0$); (d) phase and (e) intensity of the PSF calculated focusing a plane-wavefront; (f) demagnified image of the undulator photon-source (extended source).}
    \label{fig:simulations_LF}
\end{figure}

\begin{figure}
    \centering
    \includegraphics[width=15cm]{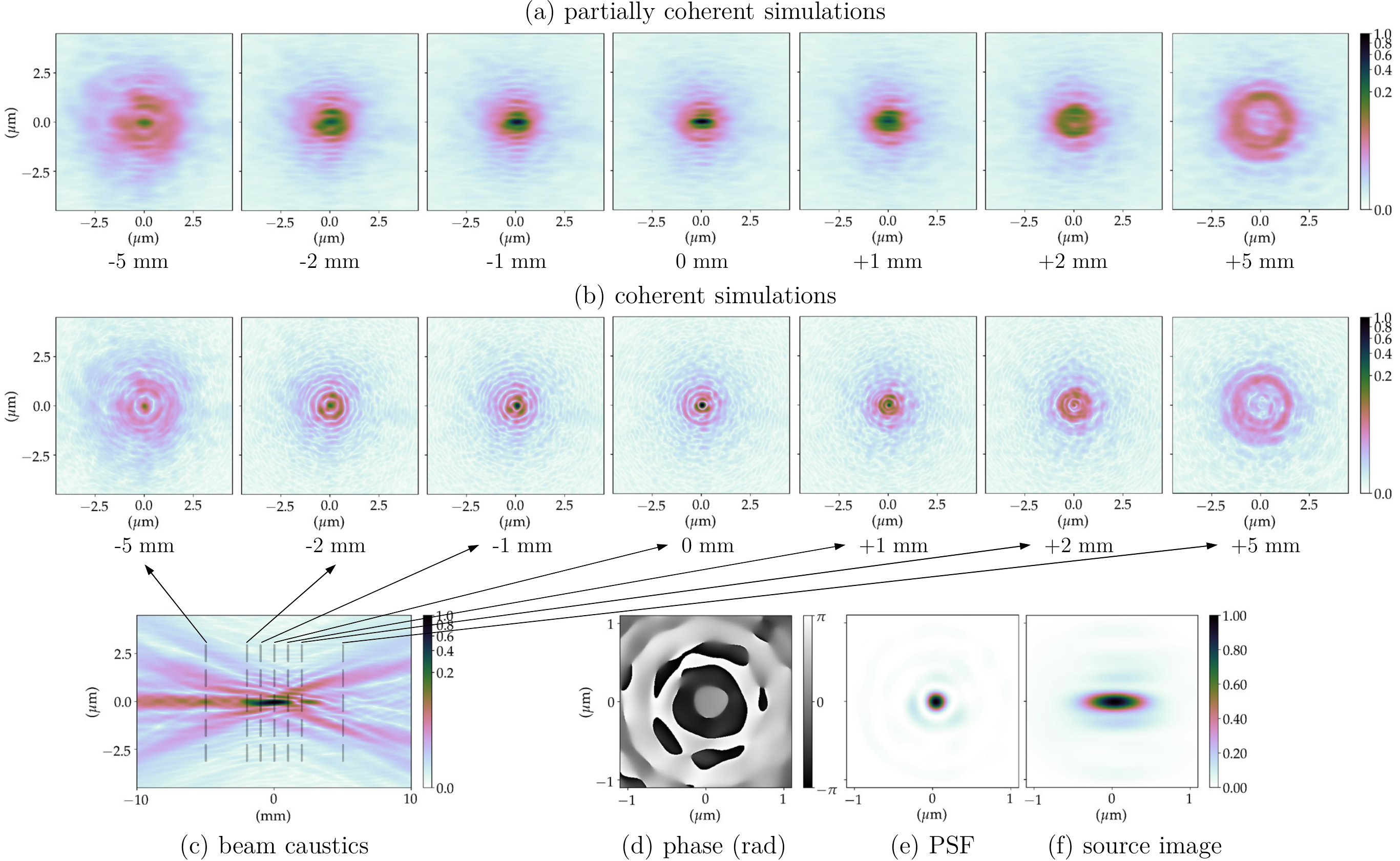} 
    \caption{CRL-MS model with the full figure errors at 8~keV. (a) partially-coherent simulations show the beam profile up- and downstream the focal position averaging 10$^{4}$ wavefronts to simulate the radiation emitted by an undulator; (b) the coherent simulations show the beam profile of a plane wavefront being focused; (c) beam propagation near the focal position (beam caustics) for a fully coherent beam (horizontal cut around $y=0$); (d) phase and (e) intensity of the PSF calculated focusing a plane-wavefront; (f) demagnified image of the undulator photon-source (extended source).}
    \label{fig:simulations_FF}
\end{figure}

The beam characteristics at the image plane are very important and the simulations show obvious differences between the CRL multi-slice with and without figure errors at that position. We complement this with investigations of the effect of optical imperfections away from the focal position, especially because several experimental applications may use a defocused beam. To get an overview of the beam evolution up- and downstream the focal position, one can propagate the wavefront along the optical axis and for each position extract a cross-section of the beam. This will be referred to as the beam caustic\footnote{Strictly speaking, the beam caustic is the envelope of light rays after passing through an optical element - see p.~60 \cite{Lawrence1972}.}. The beam cross-section for selected positions along the beam optical axis can be seen in Figs.~\ref{fig:simulations_ideal}(b)-\ref{fig:simulations_FF}(b), while Figs.~\ref{fig:simulations_ideal}(c)-\ref{fig:simulations_FF}(c) show the beam caustics for the same multi-slice CRL models. The horizontal cuts were taken at $y=0$. The zero position along the optical axis is given by the distance from the centre of the CRL to the image plane for each model (c.f. Table~\ref{tab:FocalLength}). To calculate the beam caustics, the wavefront was propagated from 10~mm upstream the focal position to 10~mm downstream in 4001 equally spaced steps along the optical axis.

\onecolumn
{}

{}

\newpage
{}

{}

\newpage
\twocolumn

\subsection{Partially coherent simulations}\label{section:partcoeh}

The PSF and beam caustic simulations shown previously are both fully-coherent calculations. They present the focusing of a perfect plane wavefront to a diffraction-limited spot. This shows the intrinsic limitations of the optical system, but inherently neglects the effects of an extended and partially coherent source.

\subsection*{X-ray source}

The emission of a single electron passing through an undulator (filament beam) is fully coherent. By changing the electron initial conditions (positions, direction and energy), propagating the emission of this electron through the beamline and adding up intensities, one can simulate partially coherent radiation if the electron beam phase space (5D) is sufficiently sampled \cite{Chubar2011}. In a conservative approach, the partially coherent simulations presented here were done using 10$^{4}$ wavefronts to ensure convergence.

For this section, we chose to implement a hypothetic beamline operating on the new Extremely Brilliant Source (ESRF-EBS) magnetic lattice \cite{orangebook}. The beamline sits on a straight section and has a CPMU18\footnote{Cryogenic Permanent Magnet Undulator with 18~mm magnetic period.} undulator as an insertion device. The undulator was tuned to its first harmonic at 8~keV for all simulations. The photon source size is $\sim71.92\times12.38~\mu\text{m}^2$ and its divergence $\sim17.66\times14.72~\mu\text{rad}^2$ (FWHM, horizontal vs. vertical). The first optical element was placed 60~m downstream of the centre of the undulator to ensure a beam footprint larger than the geometric aperture of the CRL being studied ($A_{\diameter}\sim440~\mu\text{m}$) and a constant intensity over it. The transverse coherence length at the optical system is estimated to be $\sim60\times448~\mu\text{m}^2$, from van-Cittert-Zernike theorem. If there is no cropping of the beam (e.g. use of slits to generate a secondary source), the horizontal direction is less coherent than in the vertical, leading to stronger blurring of the image in the less coherent direction - cf. \S7.5 in \cite{Goodman2017}. 

\subsection*{Beam characteristics at the focal position}\label{section:partcoeh_beamshape}

\begin{figure}
    \centering
    \includegraphics[width=7.5cm]{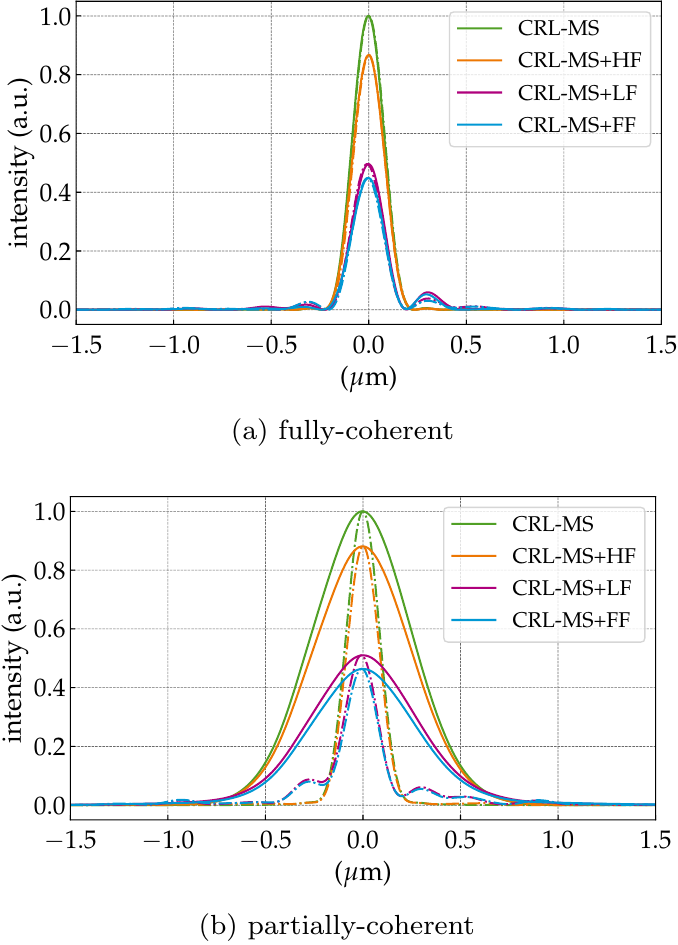} 
    \caption{ Graphical representation of the Strehl ratio. ($-$) horizontal and (-~-) vertical intensity cuts at the focal position from several CRL models under (a) fully- and (b) partially-coherent illumination. The partially-coherent simulations were done averaging the intensity of 10$^{4}$ wavefronts.}
    \label{fig:MEsimulations}
\end{figure}

The image of the extended X-ray source is similar to convolution between the geometrically demagnified image of the source and the 2D-PSF of the imaging system, provided the beam is not strongly cropped anywhere in the beamline being simulated. Figs.~\ref{fig:simulations_ideal}(e)-\ref{fig:simulations_FF}(e) show the normalised demagnified image of the undulator photon source while Table~\ref{tab:beamsizes} presents the horizontal and vertical FWHM for those simulations. Normalising the images to their peak intensity aids qualitative comparison, but omits the fact that the introduction of aberrations to the system contributes to the reduction of the peak intensity and increases the background radiation - which has been discussed in \S\ref{section:Strehl}$-$\nameref{section:Strehl}. Figure~\ref{fig:MEsimulations}(b) shows horizontal and vertical intensity cuts for the multi-slice CRL models. This is a graphical representation of the Strehl irradiance ratio, as the intensities of the aberrated models are normalised to the peak intensity of the aberration-free CRL model.

{}

\begin{table}[]
\label{tab:Strehl}
\caption{Strehl ratio calculated for the multi-slicing models (ideal and with aberrations) using the accumulated figure errors ($\sigma_z$) and Eqs.~\ref{eq:Strehl}-\ref{eq:Mahajan}. Non-physical values omitted. Partially coherent simulations were done with with 10$^{4}$ wavefronts.}
\centering
\begin{tabular}{rcccccc}
\textbf{lens model}     &\textbf{$\sigma_z$~($\mu\text{m}$)} &\textbf{S$_{\text{ratio a}}$} &\textbf{S$_{\text{ratio b}}$} &\textbf{S$_{\text{ratio c}}$} &\textbf{coherent}  &\textbf{partially-coh.}\\ \cline{1-7}
CRL-MS              & 0.00 &1.00    &1.00    &1.00    &1.00(9) &1.00(4) \\
CRL-MS+HF           & 1.77 & 0.85(5)&0.86(0) &0.86(5) &0.87(6) &0.88(1)\\
CRL-MS+LF           & 4.91 &  -     &0.19(4) &0.28(2) &0.50(1) &0.51(0)\\
CRL-MS+FF          & 5.22  & -      &0.13(5) &0.32(7) &0.45(3) &0.46(4)\\
\end{tabular}
\end{table}{}

\subsection*{Beam profile evolution along the optical axis}\label{section:partcoeh_caustic}

Calculating the full beam caustic with partially-coherent simulations is impractical using current simulation methods and computers/clusters especially if: \textit{i})  the beamline does not present a very high degree of coherence, thus requiring a very large number of wavefronts to accurately simulate the partial-coherence; \textit{ii}) the beamline has low transmission (strong beam cropping, diffraction orders outside apertures); or \textit{iii}) the sampling along the optical axis is high. Still, many applications require to operate up- or downstream of the focal position and assessing the beam quality on such positions is essential. Figures~\ref{fig:simulations_ideal}(a)-\ref{fig:simulations_FF}(a) show the beam profile evolution spanning 10~mm along the optical axis for selected positions up- and downstream the image plane. Images are displayed showing their relative intensity to the beam in the focal plane. The positions chosen were the same as in  Figs.~\ref{fig:simulations_ideal}(b)-\ref{fig:simulations_FF}(b), selected cuts along the beam caustics, so direct comparison between fully- and partially-coherent simulations can be done.
\newline

\section{Discussion}\label{section:Discussion}

In this section, we discuss the main results drawn from the simulations presented previously. We start by making considerations on the effect of optical imperfections on a (partially) coherent X-ray beam. The merit of using the Strehl ratio for X-ray lenses tolerancing is discussed. Finally, some comments on the simulation times of the several models and simulations are done.

\subsection*{The effect of optical imperfections}\label{section:discussion_imperfections}

Applying the Mar\'echal criterion (Eq.~\ref{eq:MarechalCriterion}) calculated for Beryllium lenses illuminated at 8~keV requires the accumulated projected figure errors to be $\sigma_z\leq2.08~\mu\text{m}$. Table~\ref{tab:LensSpecs} and Fig.~\ref{fig:phase} show that except for the high-frequency range, the system is operating far from ideal as the system exceeds the limit imposed by the Mar\'echal criterion.

The addition of the mid- and high-spatial frequency errors ($\sigma_z\sim1.77~\mu\text{m}$) are related to scattering around the focused beam, contributing thus to increased background and consequently reducing the peak intensity following \citename{Harvey1995a}~\citeyear{Harvey1995a}. Using a linear scale, both the ideal PSF and the demagnified source image in Figs.~\ref{fig:simulations_ideal}(e)-(f) are indistinguishable from their aberrated counterparts in Figs.~\ref{fig:simulations_HF}(e)-(f), which is due to the fact that the accumulated figure error complies to the Mar\'echal criterion.  When considering the low-spatial-frequency figure errors ($\sigma_z\sim4.91~\mu\text{m}$), however, concentric faint rings start appearing on the PSF. Homogeneous concentric rings on the PSF are a classical signature of spherical aberration, which is a major component of the LF figure errors - cf. $Z_{11}$ in Fig.~\ref{fig:phase}(g). The predominance of spherical aberration on 2D parabolic Be lenses has already been observed; see Fig.~ 6.14 of \citename{Seiboth2016b}~\citeyear{Seiboth2016b}. The PSF due to spherical aberration can be seen also in Figs.~8.5 and 8.6 from \cite{Mahajan2011}. In the partially-coherent simulations, the rings around the main lobe seen at the PSF simulations are stretched horizontally to the point that their visibility is maintained vertically, but horizontal cuts (Fig.~\ref{fig:MEsimulations}(b)) show almost no trace of them, due to the reduction in transverse horizontal coherence (blurring effect). Small misalignments between the lenslets and some residual tilt from the LF errors contribute to a lateral displacement of the beam in the image plane - this is observed also in Figs.~\ref{fig:simulations_LF}(e) and \ref{fig:simulations_FF}(e). Using the full-frequency-range figure errors ($\sigma_z\sim5.22~\mu\text{m}$) yields a combined effect that is analogous to the superposition of the HF and LF figure errors. The CRL-FF model can be seen in Fig.~\ref{fig:simulations_FF}. The diffraction effects from the aperture of the CRL are not easily observable because the system has an apodised Gaussian intensity at the exit pupil \cite{Mahajan1986}.

The addition of figure errors change the beam profile more significantly  up- and downstream of the focal position. Fig.~\ref{fig:simulations_ideal}(a)-(c) show the focusing done by the multi-slicing model of the CRL without any optical imperfections. Introducing the HF errors does not significantly change the beam shape as they contribute to the scattering of light outside the beam envelope defined by the beam caustics - cf. Figs.~\ref{fig:simulations_HF}(a)-(c). The LF errors act to change the beam shape dramatically as can be seen in Figs.~\ref{fig:simulations_LF}(a)-(c) and Figs.~\ref{fig:simulations_FF}(a)-(c). Upstream of the image plane, a persistant central lobe is observed, albeit much less intense, with a high background around it thus reducing the signal to noise ratio. Downstream, the beam has a drop in intensity in the middle, looking like a doughnut when a cut transverse the optical axis is made. This behaviour is observed both on fully- and partially-coherent simulations. Such beam caustics have been extensively reported by experimental groups working under high coherent conditions, with similar optics and ptychographic reconstruction of X-ray beams - cf. Fig.~3 in \cite{Schropp2013}, Fig.~2 in \cite{Seiboth2016} and Fig.~3 in \cite{Gasilov2017}.

\begin{figure}
    \centering
    \includegraphics[width=7.5cm]{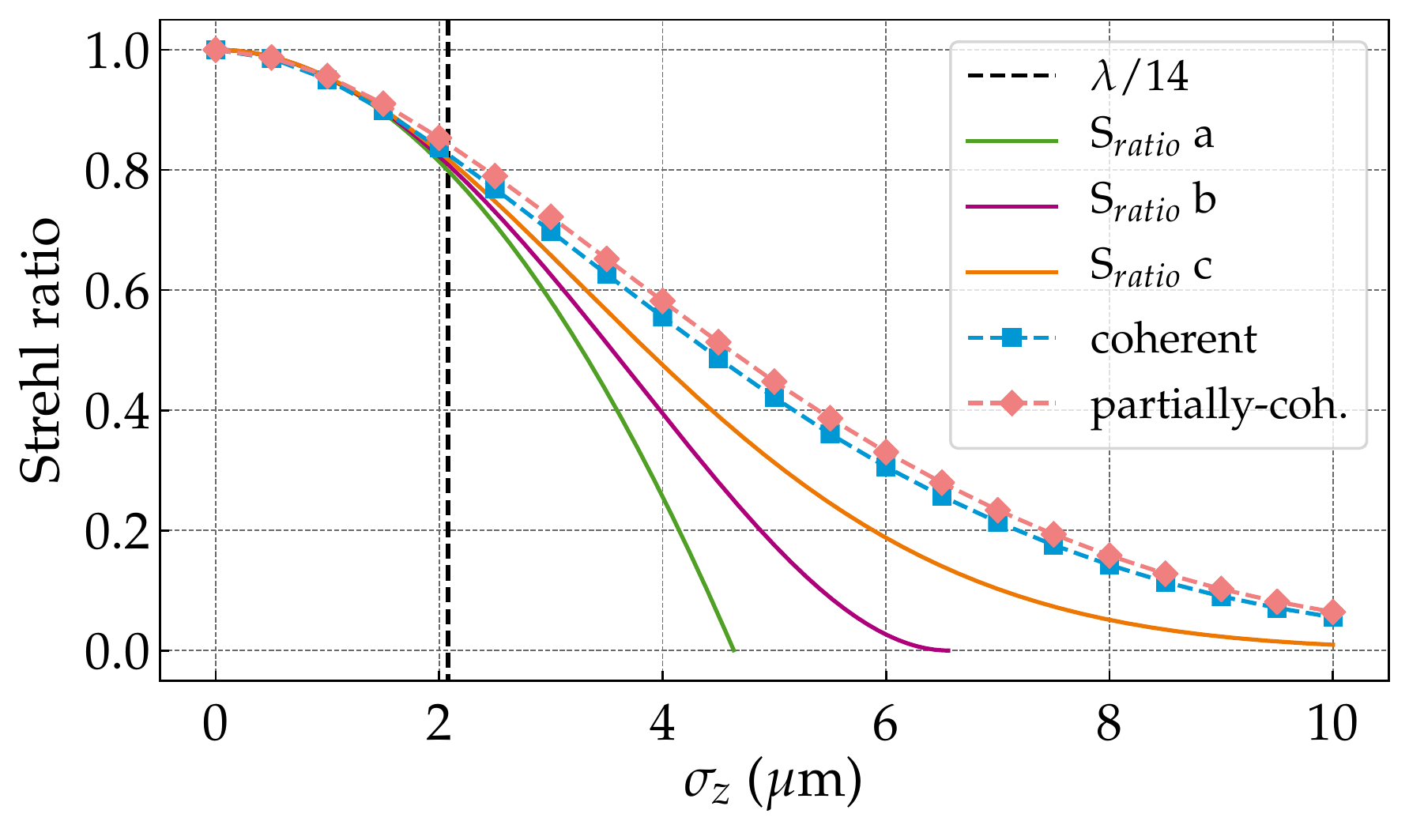} 
    \caption{Strehl ratio from numerical simulations and from the application of different approximations (Eqs.~\ref{eq:Strehl}-\ref{eq:Mahajan}) as a function of the figure error $\sigma_z$ from a lens stack made of Beryllium illuminated at 8~keV. The vertical dashed black line indicates the maximum tolerable thickness (Eq.~\ref{eq:ThickLim}) for complying with the Mar\'echal criterion (Eq.~\ref{eq:MarechalCriterion}), that is, $\sigma_{\lambda/14}\approx2.08~\mu\text{m}$. The partially-coherent simulations were done with 10$^{4}$ wavefronts.}
    \label{fig:Strehl}
\end{figure}{}

\subsection*{The Strehl ratio for X-ray lenses}\label{section:discussion_strehl}

The Strehl ratio for the CRL models is presented in Table~\ref{tab:Strehl}. In the numerical simulations, the intensity at the centre of the beam is normalised to the intensity obtained by a single-lens equivalent system. Due to this fact, the CRL-MS model has slightly more intensity in the central peak than the single-lens equivalent ($\sim0.9\%$ and $\sim0.4\%$ for the coherent and partially-coherent cases, respectively), which is explained by the fact that the X-rays are continuously being focused inside the CRL\footnote{This effect is negligible for a short CRL, but it can become important when the number of elements is drastically increased \cite{Schroer2005}.} - cf. \cite{Schroer2005}.

Our results show (Table~\ref{tab:Strehl}) that applying the Strehl ratio calculated from analytic equations (Eqs.~\ref{eq:Strehl}-\ref{eq:Mahajan}) overstates the effect of moderate figure errors on the overall system performance. In order to understand the dependency of the numerically simulated Strehl ratio versus height error r.m.s. we used the CRL-MS+HF model and scaled each individual figure error (Table~\ref{tab:LensSpecs}) by a constant value to allow for a scanning of the total projected figure error $\sigma_z$. The results in Figure~\ref{fig:Strehl} show the expected Strehl ratio as a function of the projected figure errors $\sigma_z$ for different analytical approximations (Eqs.~\ref{eq:Strehl}-\ref{eq:Mahajan}) and for the numerical calculations with a fully- and partially-coherent illumination. All approaches show very good agreement up to $S_\text{ratio}>0.8$, when they start diverging. The expressions for $S_\text{ratio a}$ (Eq.~\ref{eq:Strehl}) and  $S_\text{ratio b}$ (Eq.~\ref{eq:Marechal}) can be considered as approximations for $S_\text{ratio c}$ (Eq.~\ref{eq:Mahajan}), therefore are only expected to be valid over a restricted range (large $S_\text{ratio}$). A fit of the simulation data  (coral rhombuses and blue squares in Fig.~\ref{fig:Strehl}) give:
\begin{subequations}\label{eq:Strhel_Exp_pre}
\begin{align}
    S_{\text{ratio coh.}}&\approx\exp{\big(-2.32\cdot10^{10}\sigma_z^2 -6.13\cdot10^4\sigma_z + 2.54\cdot10^{-2}\big)},\\
    S_{\text{ratio part.-coh.}}&\approx\exp{\big(-2.28\cdot10^{10}\sigma_z^2 -5.07\cdot10^4\sigma_z + 2.29\cdot10^{-2}\big)}.
\end{align}
\end{subequations}
Unfortunately, due to the nature of the projected figure errors (in the range of few micrometres r.m.s.), we are not able to discard the non-quadratic terms. We can rewrite Eqs.~\ref{eq:Strhel_Exp_pre} as:
\begin{align}\label{eq:Strhel_Exp}
    S_{\text{ratio simulation}}&\approx\exp{\bigg[- \bigg(\frac{2\pi}{\lambda}\bigg)^2\big(\kappa_1\Delta\Phi\big)^2-\frac{2\pi}{\lambda}\kappa_2\Delta\Phi-\kappa_3\bigg]},
\end{align}
where $\kappa$ are scaling constants that, in principle, the number of elements, lens material, energy and, mostly importantly, the spatial distribution of the accumulated errors over the optical element aperture. For our particular examples, $\kappa_1=0.71$, $\kappa_2=0.28$ and $\kappa_3=2.54\cdot10^{-2}$ for the coherent case and $\kappa_1=0.70$, $\kappa_2=0.24$ and $\kappa_3=2.29\cdot10^{-2}$ for the partially-coherent case. When comparing Eq.~\ref{eq:Strhel_Exp} with Eq.~\ref{eq:Mahajan}, a $\kappa<1$ suggests that there is some weighting of the phase errors reducing their effect, but simply multiplying the accumulated phase errors (cf. Table~\ref{tab:LensSpecs}) with the normalised optical system transmission (cf. Fig.~\ref{fig:EffectiveAperure}) does not allow prediction of $\kappa$ and we leave this as an open question at the time of writing.

Following the recent discussion about the pertinence of the $S_\text{ratio}\geq0.8$ as an indicator of optical quality for the X-ray regime and the performance of such optical systems away from the focal position \cite{Cocco2015, Cocco2019}, we can observe from our simulations (Figs.~\ref{fig:simulations_ideal}-\ref{fig:simulations_FF}) that in terms of wavefront preservation, X-ray lenses are apparently more susceptible to the low-frequency figure errors, as they are the ones that change the beam profile up- and downstream the focal position. The high-frequency errors lead to scattering of the beam and speckles, but generally, do not change the beam shape even away from the focal position. It is clear from Figs.~\ref{fig:simulations_LF} and \ref{fig:simulations_FF} that the Strehl ratio encountered at the focal position (cf. Fig.~\ref{fig:MEsimulations}) is not preserved up- or downstream from it. Fortunately, the low frequencies are those which can be readily corrected by the fitting of corrective optics \cite{Seiboth2017}. Corrective plates aim at increasing the Strehl ratio in the low-frequency range, leaving the high-frequencies as the bottleneck for corrected systems performance.

\subsection*{Simulation time}\label{section:discussion_time}

Increasing the complexity of the simulation model comes at the expense of increasing the overall simulation time, but as long as the transverse wavefront sampling is maintained, memory consumption is not affected from one model to another. The time increase in the simulations is mainly due to: \textit{i}) increase in the number of drift spaces and the number of optical elements; \textit{ii}) from reading the densely sampled metrology data. Table~\ref{tab:simulation_time} presents the typical simulation times for this work. Those are particularly high because the transverse sampling of the wavefronts is several times larger than the nominal minimal sampling necessary to mitigate artefacts or under-resolved features on the wavefront. Employing 10$^{4}$ wavefronts for the partially coherent simulations is also exaggerated, but was done to ensure that any changes on the simulation come from the change of model being studied and not from statistical nature of the sampling of the electron-beam phase-space. The simulation times presented on Table~\ref{tab:simulation_time} can be certainly be reduced without loss of accuracy by adopting a more sensible sampling. 

\begin{table}[]
\label{tab:simulation_time}
\caption{Summary of the simulation times for different CRL models. From the most simple one (single lens equivalent) up to the more complex multi-slicing (MS) with figure errors. Simulations were performed on a Intel(R) Xeon(R) CPU E5-2680 v4 @ 2.40GHz cluster at the ESRF. Partially coherent calculations were done using 28 cores in parallel.}
\centering
\begin{tabular}{rccc}
\multicolumn{1}{c}{\textbf{model}} & \textbf{fully coherent} &\textbf{ partially coherent} & \textbf{caustics}\\\cline{1-4}
single lens equiv.                 &33~s            &2~h~44~min  &1~h~32~min               \\
multi-slicing                      &58~s            &5~h~12~min  &1~h~33~min               \\
MS + figure errors                 &2~min~48~s      &5~h~42~min  &1~h~35~min               \\\cline{2-4}
\multicolumn{1}{c}{}               & (1 wavefront)  & (10$^{4}$ wavefronts)  & (4000 pts.)
\end{tabular}
\end{table}

\section{Conclusion}

Building on physical optics concepts and already implemented optical elements in SRW, we have expanded the concept of the complex transmission element representation of the CRL to account not only for its thick element nature but also real imperfections obtained with at-wavelength metrology. We have studied the adequacy of commonly used design equations and figures of merit by doing coherent and partially-coherent simulations. We were able to accurately simulate the effects of figure errors on beam shape and intensity along the optical axis. Our simulations of the beam caustics compare well with experimental data from other research groups using the same type of Be lenses. We show that using the Strehl ratio formulations given by Eqs.~\ref{eq:Strehl}-\ref{eq:Mahajan} leads to an underestimation of the system performance if the total projected figure errors are larger than the limit imposed by the $\lambda/14$ criterion (Mar\'echal criterion for optical quality). We see an immediate application to lens tolerancing and guidelines for accepting or not commercial optical elements and in-house lens production and testing of X-ray lenses (quality control). By decomposing the figure errors in frequency ranges, we note that the strongest contribution to the degradation of the wavefront both in focus and away from it comes from the low-frequency range, which is where corrective optics are most efficient. By being able to add individual lens profiles to a lens stack we envisage the possibility of calculating corrective optics for an arbitrary lens combination offline, as opposed to experimentally measuring the wavefront phase errors of the lens stack as proposed by \citename{Seiboth2017}~\citeyear{Seiboth2017}. 



\ack{Acknowledgements}

The authors are thankful to O. Chubar (NSLS-II/BNL) and D. Cocco (ALS/LBNL) for the helpful discussions, to C. Detlefs (ID06/ESRF) for discussions, comments on the manuscript and providing the lenses for metrology and to R. Cojucaro (BM05/ESRF) for helping during the experimental sessions. 

\bibliographystyle{iucr}
\referencelist{iucr}



\end{document}